# COLLISION OF ELASTIC DROP WITH THIN CYLINDER

Alla O. Rudenko, Alexey I. Fedyushkin, Aleksey N. Rozhkov

*Ishlinsky Institute for Problems in Mechanics, Russian Academy of Sciences,*
*101(k.1) Prospect Vernadskogo, Moscow 119526, Russia*

## ABSTRACT

The collision of water and elastic liquid drops with a thin cylinder (thread) is studied. The droplet flight trajectory and the cylinder axis are mutually perpendicular. Attention is focused on the difference between collisions of water drops and drops of elastic fluids. In the experiments, the drop diameter was 3 mm, the diameter of horizontal stainless steel cylinders was 0.4 and 0.8 mm. The drops were formed by slowly pumping liquid through a vertical stainless steel capillary with an outer diameter of 0.8 mm, from which droplets were periodically detached under the action of gravity. The droplet velocity before collision was defined by the distance between the capillary cut and the target (cylinder); in experiments, this distance was approximately 5, 10, and 20 mm. The drop velocities before impact are estimated in the range of 0.2–0.5 m/s. The collision process was monitored by high-speed video recording methods with a frame rate of 240 and 960 Hz. The test liquids were water and aqueous solutions of polyacrylamide of molecular weight 11 million and concentrations of 100 and 1000 ppm (PAM-100 and PAM-1k). Experiments have shown that, depending on the drop impact height and polymer concentration, different scenarios of a drop collision with a thin cylinder are possible: a short-term recoil of a drop from an obstacle, a drop flowing around a cylindrical obstacle while maintaining the continuity of the drop, the breakup of a drop into two secondary drops, one of which can continue flight and the other one is captured by the cylinder, or both secondary droplets continue to fly, the drop also can be captured by the cylinder, until the impact of the next drop(s) forces the accumulated drop detach from the cylinder. Numerical modeling satisfactorily reproduces the phenomena observed in the experiment.

## 1. Introduction

Protection against infections transmitted by airborne droplets is carried out through the use of medical masks and filters that either inhibit or retard the movement of droplets, possible carriers of infections. Pathogenic droplets are formed by breathing, talking, coughing, sneezing of a sick person and enter the body of a healthy person by inhaling air containing such droplets. The work is aimed at studying the mechanisms of collision of liquid droplets with the material of the masks and filters. An elementary act of such an interaction is simulated experimentally, namely, the fall of a drop onto the lateral surface of a thin cylinder imitating the fibrous component of the mask. Unlike previous experiments [1], not only drops of Newtonian fluids, but also viscoelastic fluids, the rheology of which corresponds to real oral and bronchial fluids [2], participate in collisions.

## 2. Materials and methods

The tests were carried out on aqueous solutions of polyacrylamide with a molecular weight of 11 million with a mass fraction of 0, 100 and 1000 ppm (hereinafter referred to as Water, PAM-100 and PAM-1k). The results of the rheological test of liquids are presented in [3].

The schematic of the experiment and experimental setup are shown in Fig. 1.

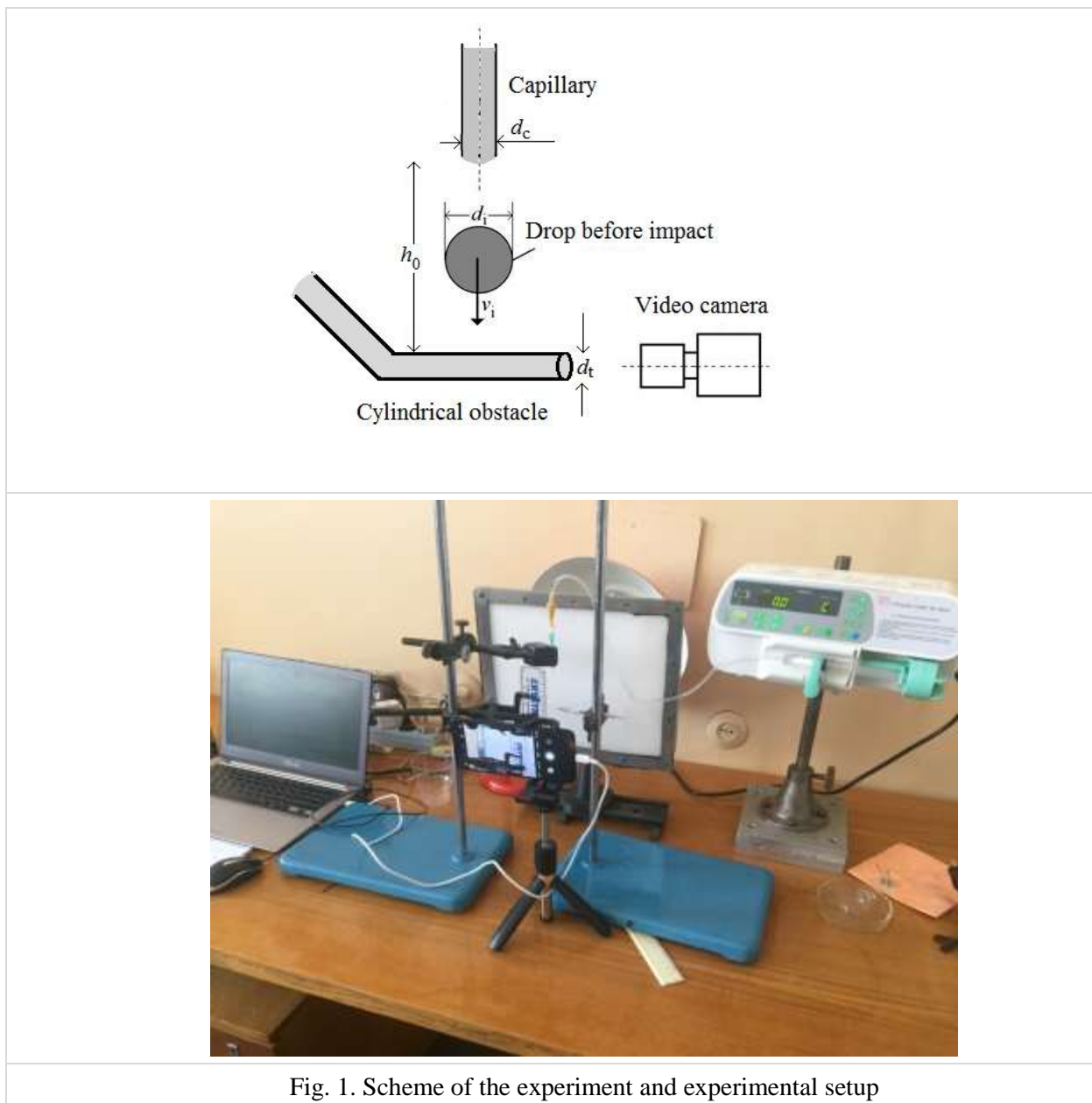

Fig. 1. Scheme of the experiment and experimental setup

A drop of liquid was formed at the end of a capillary, the function of which was performed by an injection needle with outer diameter of $d_c = 0.8$ mm. The liquid slowly filled the drop using a SINO SN-50C6 infusion pump. Upon reaching a certain weight, the drop was detached from the capillary [2]. Then drop collided with a transverse cylindrical obstacle. Another injection needle with a diameter of $d_t = 0.4$ or $0.8$ mm used as an obstacle. To increase the contrast and remove foreign objects from the field of photography, the needle was bent approximately at a right angle, so that only the drop and the tip of the needle are visible in the frames (Fig. 2). In the experiments, the distance from the cut of the upper needle to the lateral surface of the lower one was $h_0 \approx 5$, 10, and 20 mm. The collision process was monitored using video recording with a frame rate of $f = 240$ and $960$ Hz. The smartphones iPhone6 (240 Hz) and Honor 30S (CDY-NX9A) (960 Hz) were used for video recording. The processing of video frames made it possible to determine the diameter $d_i \approx 3$ mm and estimate the velocity $v_i \approx 0.2-0.5$ m/s of the droplet before contact with the obstacle, as well as to trace all the phases of the collision shown in Fig. 2−19.

## 3. Results

### 3.1 Experimental observations

The observed processes and features of the collision are presented in Table 1.

Table 1

| | $h_0 \approx 5$ mm | | $h_0 \approx 10$ mm | | $h_0 \approx 20$ mm | |
|---|---|---|---|---|---|---|
| | $d_t$=0.4 mm $d_c$=0.8 mm | $d_t$=0.8 mm $d_c$=0.8 mm | $d_t$=0.4 mm $d_c$=0.8 mm | $d_t$=0.8 mm $d_c$=0.8 mm | $d_t$=0.4 mm $d_c$=0.8 mm | $d_t$=0.8 mm $d_c$=0.8 mm |
| Water | Rebound and fall. Fig. 2 | Rebound and fall. Fig.5 | Fall. Fig. 3 | Disintegration into two parts, capture of one and detachment of the other, as well as detachment of the remaining drop by the next drop. Fig. 6 | Fall. Fig. 4 | Break into two parts, coalescence and fall. Fig. 7 |
| PAM-100 | Rebound and fall. Fig.8 | Rebound, capture and detachment of the 3rd drop. Fig. 11 | Rebound, capture and fall with the next drop. Fig. 9 | Rebound, capture and fall with the next drop. Fig. 12 | Splitting into two parts, capturing one and detachment the other, and detachment by the next drop. Fig. 10 | Suspending, capture and detachment the next drop. Fig. 13 |
| PAM-1k | Capture, fall by the next drop with suspending. Fig. 14 | Capture, fall by the next drop with suspending. Fig. 17 | Capture, fall by the next drop with suspending. Fig. 15 | Capture, fall by the next drop with suspending. Fig. 18 | Rebound, capture, fall by the next drop with suspending. Fig. 16 | Capture, fall by the 3rd drop with suspending. Fig. 19 |

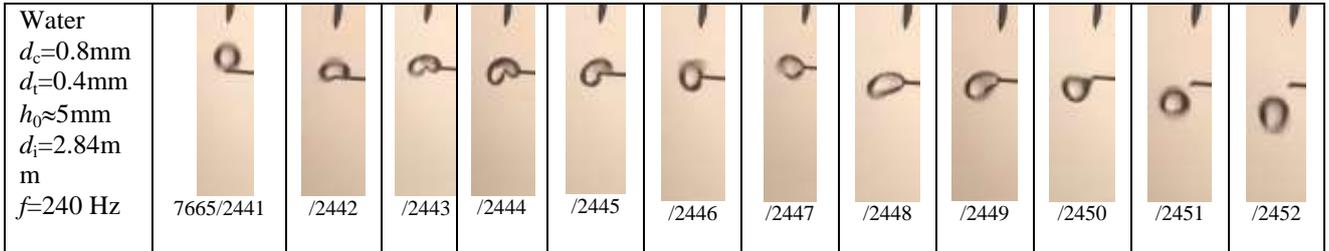

Water
$d_c$=0.8mm
$d_t$=0.4mm
$h_0$≈5mm
$d_i$=2.84mm
$f$=240 Hz

7665/2441 /2442 /2443 /2444 /2445 /2446 /2447 /2448 /2449 /2450 /2451 /2452

Fig. 2. An example of rebound and fall of a water drop from an obstacle at $d_c$ = 0.8mm, $d_t$ = 0.4mm, $h_0$≈5mm. The numbers under the frames are the number of the video and the numbers of the frames in the video

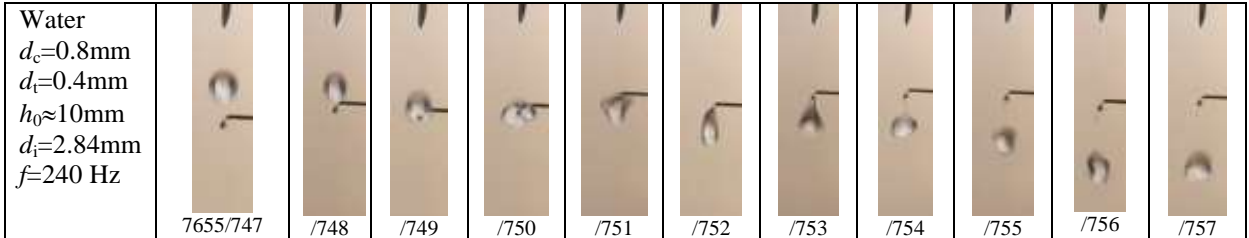

Water
$d_c$=0.8mm
$d_t$=0.4mm
$h_0$≈10mm
$d_i$=2.84mm
$f$=240 Hz

7655/747 /748 /749 /750 /751 /752 /753 /754 /755 /756 /757

Fig. 3. An example of detachment of a water drop from an obstacle at $d_c$=0.8mm, $d_t$=0.4mm, $h_0$≈10mm. The numbers under the frames are the number of the video and the numbers of the frames in the video

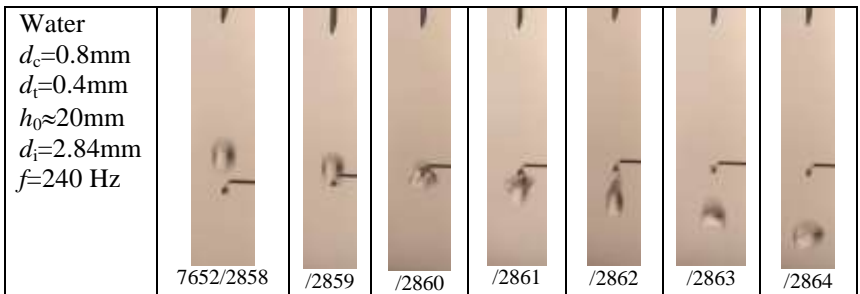

Water
$d_c$=0.8mm
$d_t$=0.4mm
$h_0$≈20mm
$d_i$=2.84mm
$f$=240 Hz

7652/2858 /2859 /2860 /2861 /2862 /2863 /2864

Fig. 4. An example of detachment of a water drop from an obstacle at $d_c$=0.8mm, $d_t$=0.4mm, $h_0$≈20mm. The numbers under the frames are the number of the video and the numbers of the frames in the video

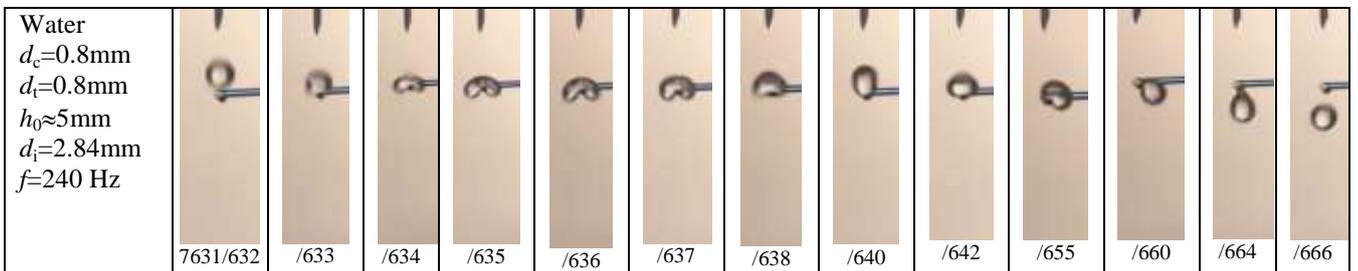

Water
$d_c$=0.8mm
$d_t$=0.8mm
$h_0$≈5mm
$d_i$=2.84mm
$f$=240 Hz

7631/632 /633 /634 /635 /636 /637 /638 /640 /642 /655 /660 /664 /666

Fig. 5. An example of a rebound and detachment of a water drop from an obstacle when $d_c$=0.8mm, $d_t$=0.8mm, $h_0$≈5mm. The numbers under the frames are the number of the video and the numbers of the frames in the video

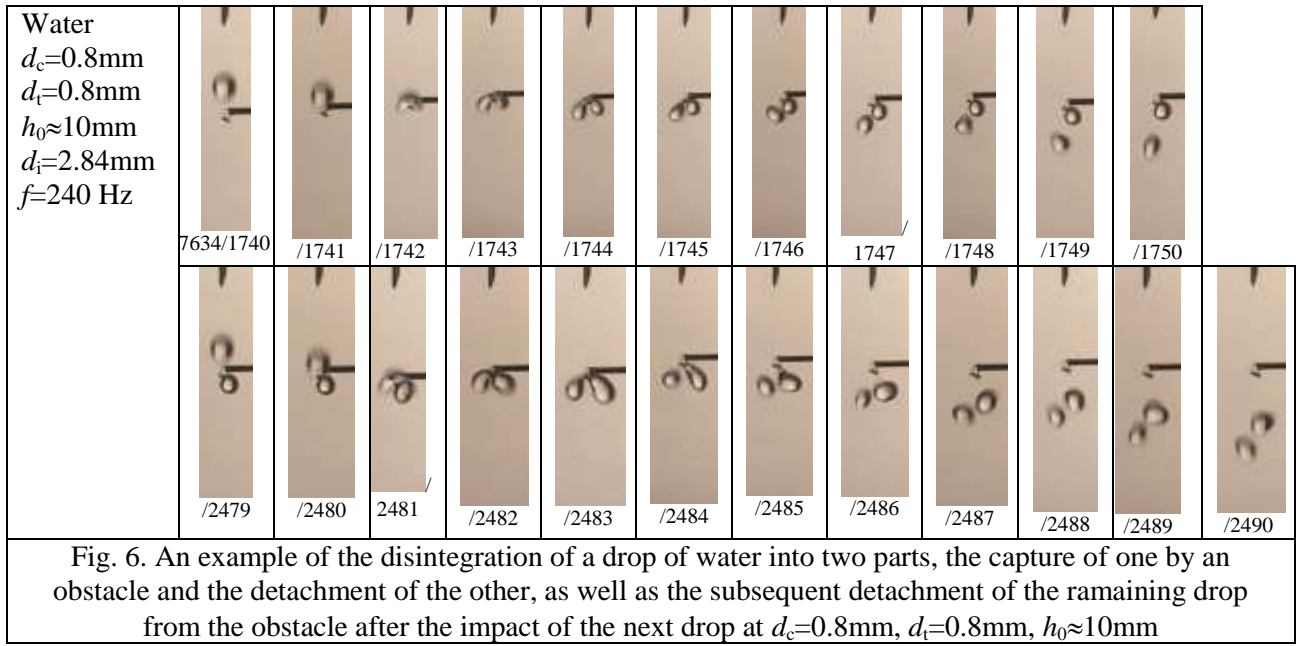

Water
$d_c$=0.8mm
$d_t$=0.8mm
$h_0$≈10mm
$d_i$=2.84mm
$f$=240 Hz

7634/1740 /1741 /1742 /1743 /1744 /1745 /1746 1747 /1748 /1749 /1750

/2479 /2480 2481 /2482 /2483 /2484 /2485 /2486 /2487 /2488 /2489 /2490

Fig. 6. An example of the disintegration of a drop of water into two parts, the capture of one by an obstacle and the detachment of the other, as well as the subsequent detachment of the ramaining drop from the obstacle after the impact of the next drop at $d_c$=0.8mm, $d_t$=0.8mm, $h_0$≈10mm

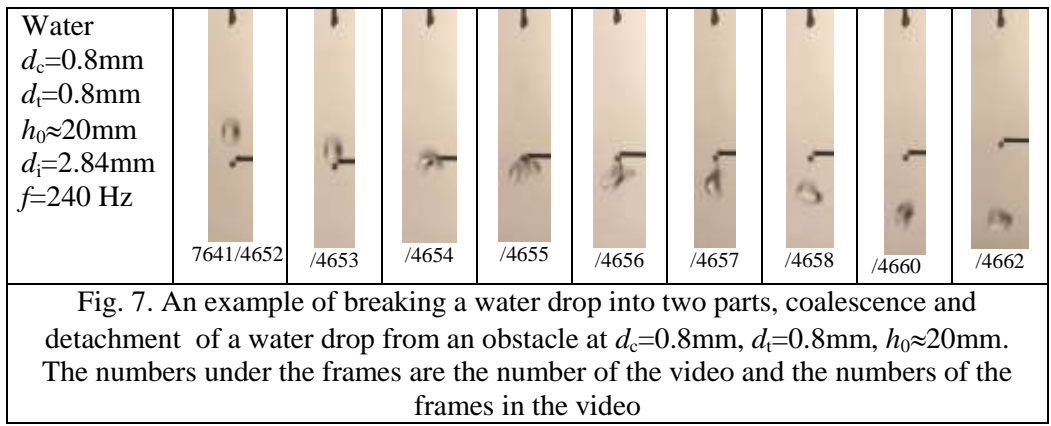

Water
$d_c$=0.8mm
$d_t$=0.8mm
$h_0$≈20mm
$d_i$=2.84mm
$f$=240 Hz

7641/4652 /4653 /4654 /4655 /4656 /4657 /4658 /4660 /4662

Fig. 7. An example of breaking a water drop into two parts, coalescence and detachment of a water drop from an obstacle at $d_c$=0.8mm, $d_t$=0.8mm, $h_0$≈20mm. The numbers under the frames are the number of the video and the numbers of the frames in the video

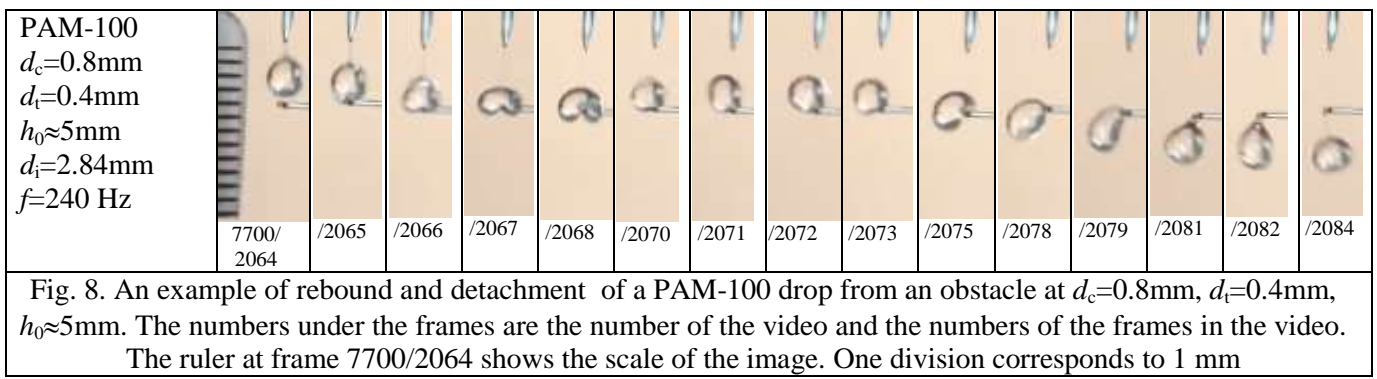

PAM-100
$d_c$=0.8mm
$d_t$=0.4mm
$h_0$≈5mm
$d_i$=2.84mm
$f$=240 Hz

7700/2064 /2065 /2066 /2067 /2068 /2070 /2071 /2072 /2073 /2075 /2078 /2079 /2081 /2082 /2084

Fig. 8. An example of rebound and detachment of a PAM-100 drop from an obstacle at $d_c$=0.8mm, $d_t$=0.4mm, $h_0$≈5mm. The numbers under the frames are the number of the video and the numbers of the frames in the video. The ruler at frame 7700/2064 shows the scale of the image. One division corresponds to 1 mm

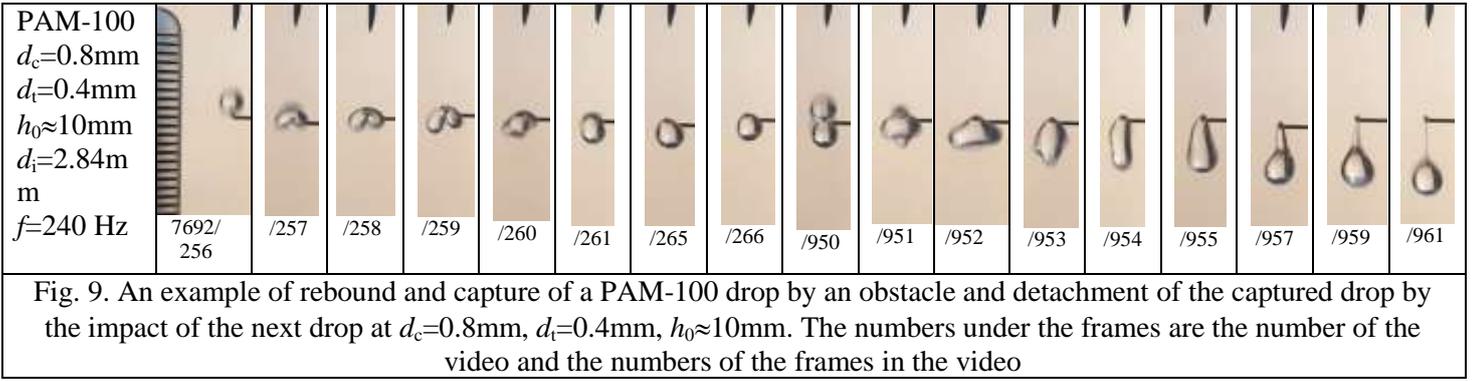

PAM-100
$d_c$=0.8mm
$d_t$=0.4mm
$h_0$≈10mm
$d_i$=2.84mm
$f$=240 Hz

Fig. 9. An example of rebound and capture of a PAM-100 drop by an obstacle and detachment of the captured drop by the impact of the next drop at $d_c$=0.8mm, $d_t$=0.4mm, $h_0$≈10mm. The numbers under the frames are the number of the video and the numbers of the frames in the video

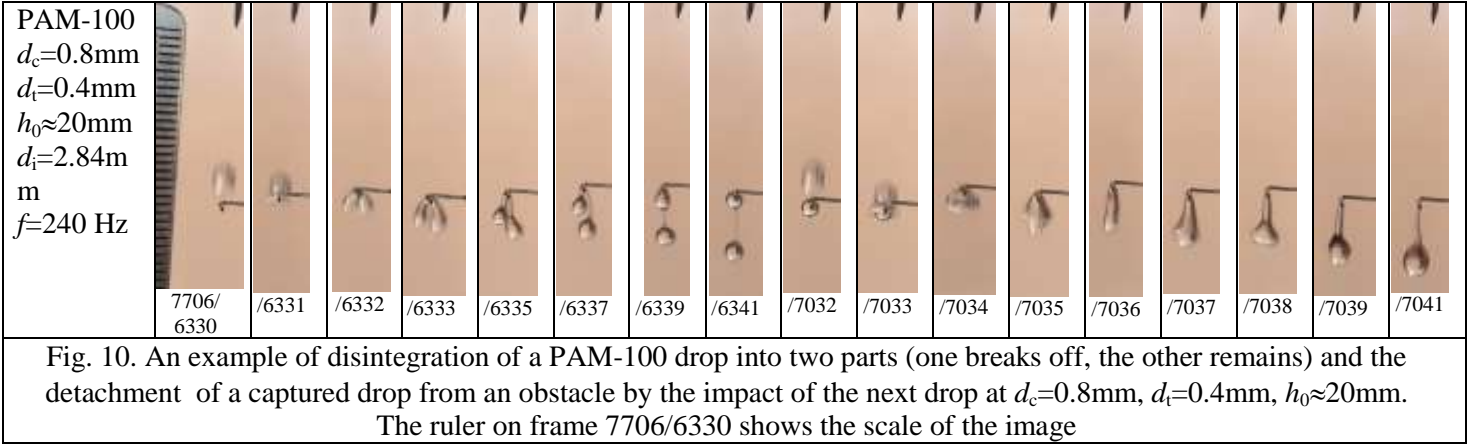

PAM-100
$d_c$=0.8mm
$d_t$=0.4mm
$h_0$≈20mm
$d_i$=2.84mm
$f$=240 Hz

Fig. 10. An example of disintegration of a PAM-100 drop into two parts (one breaks off, the other remains) and the detachment of a captured drop from an obstacle by the impact of the next drop at $d_c$=0.8mm, $d_t$=0.4mm, $h_0$≈20mm. The ruler on frame 7706/6330 shows the scale of the image

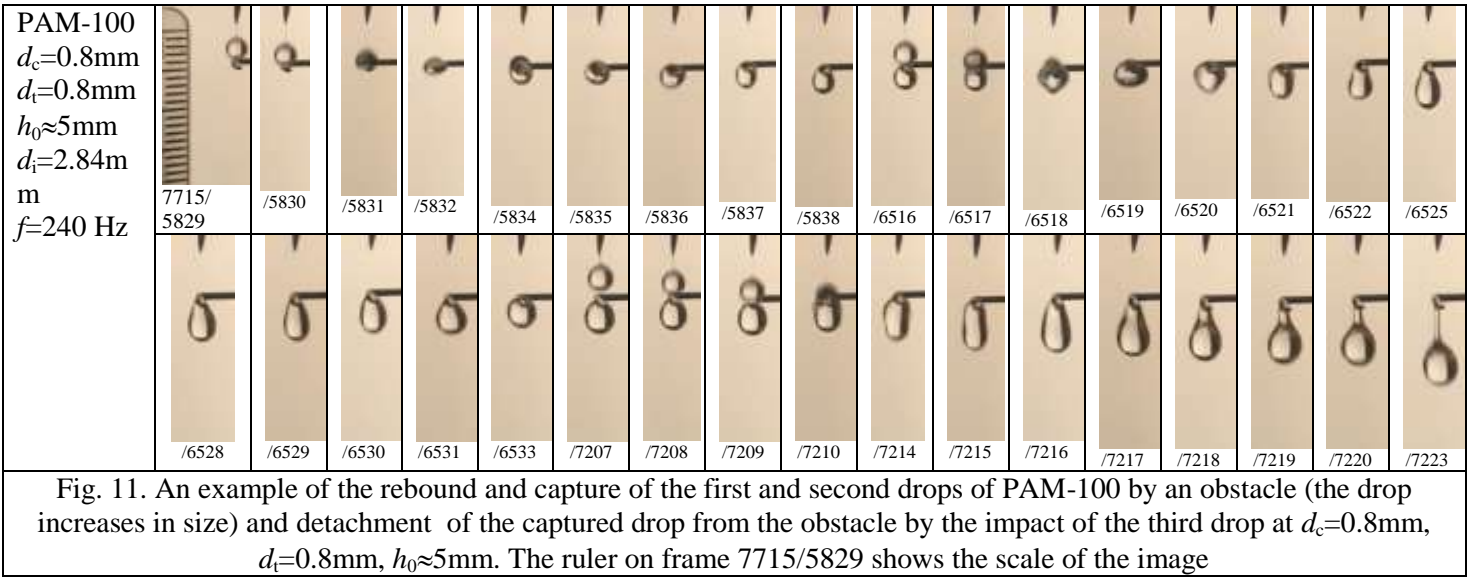

PAM-100
$d_c$=0.8mm
$d_t$=0.8mm
$h_0$≈5mm
$d_i$=2.84mm
$f$=240 Hz

Fig. 11. An example of the rebound and capture of the first and second drops of PAM-100 by an obstacle (the drop increases in size) and detachment of the captured drop from the obstacle by the impact of the third drop at $d_c$=0.8mm, $d_t$=0.8mm, $h_0$≈5mm. The ruler on frame 7715/5829 shows the scale of the image

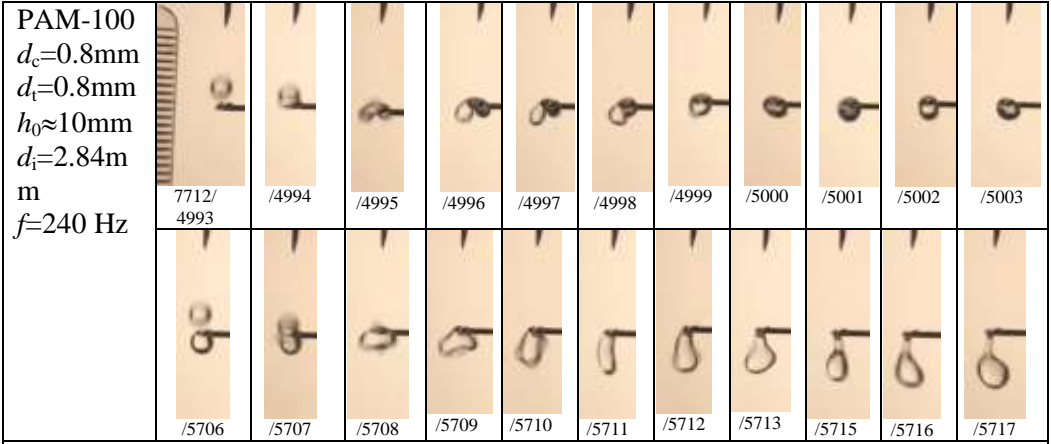

PAM-100
$d_c$=0.8mm
$d_t$=0.8mm
$h_0$≈10mm
$d_i$=2.84mm
$f$=240 Hz

7712/4993 /4994 /4995 /4996 /4997 /4998 /4999 /5000 /5001 /5002 /5003

/5706 /5707 /5708 /5709 /5710 /5711 /5712 /5713 /5715 /5716 /5717

Fig. 12. An example of a rebound, capture of a PAM-100 drop by an obstacle and detachment of the captured drop by the impact of the next drop at $d_c$=0.8mm, $d_t$=0.8mm, $h_0$≈10mm. The ruler on frame 7712/4993 shows the scale of the image

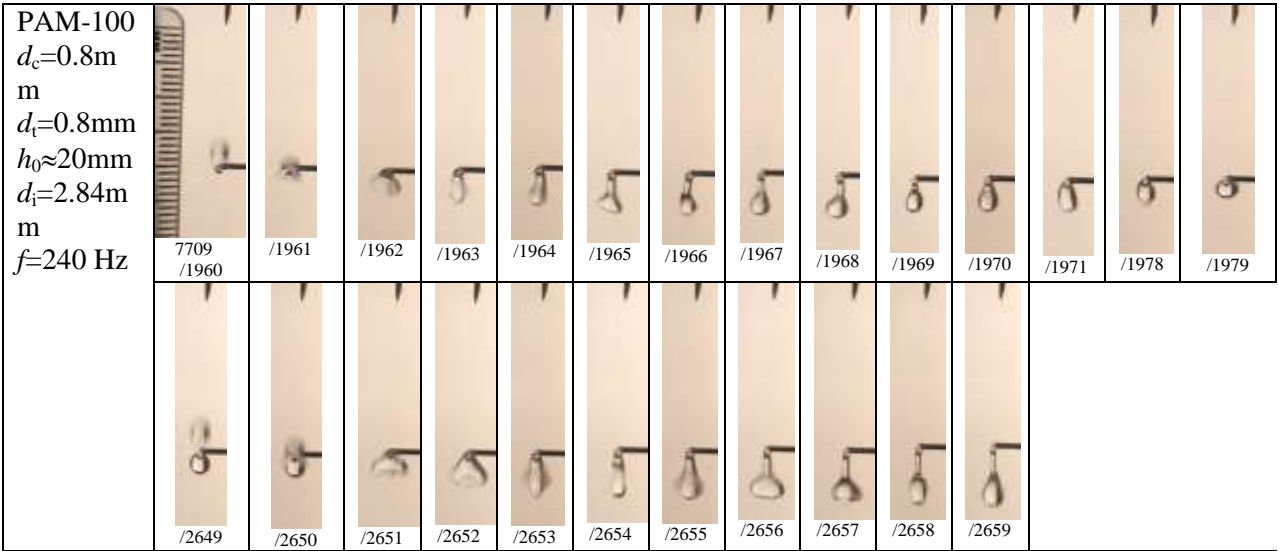

PAM-100
$d_c$=0.8mm
$d_t$=0.8mm
$h_0$≈20mm
$d_i$=2.84mm
$f$=240 Hz

7709/1960 /1961 /1962 /1963 /1964 /1965 /1966 /1967 /1968 /1969 /1970 /1971 /1978 /1979

/2649 /2650 /2651 /2652 /2653 /2654 /2655 /2656 /2657 /2658 /2659

Fig. 13. An example of sagging, capture of a PAM-100 drop by an obstacle, and detachment of the captured drop by the impact of the next drop when $d_c$=0.8mm, $d_t$=0.8mm, $h_0$≈20mm. The ruler on frame 7709/1960 shows the scale of the image

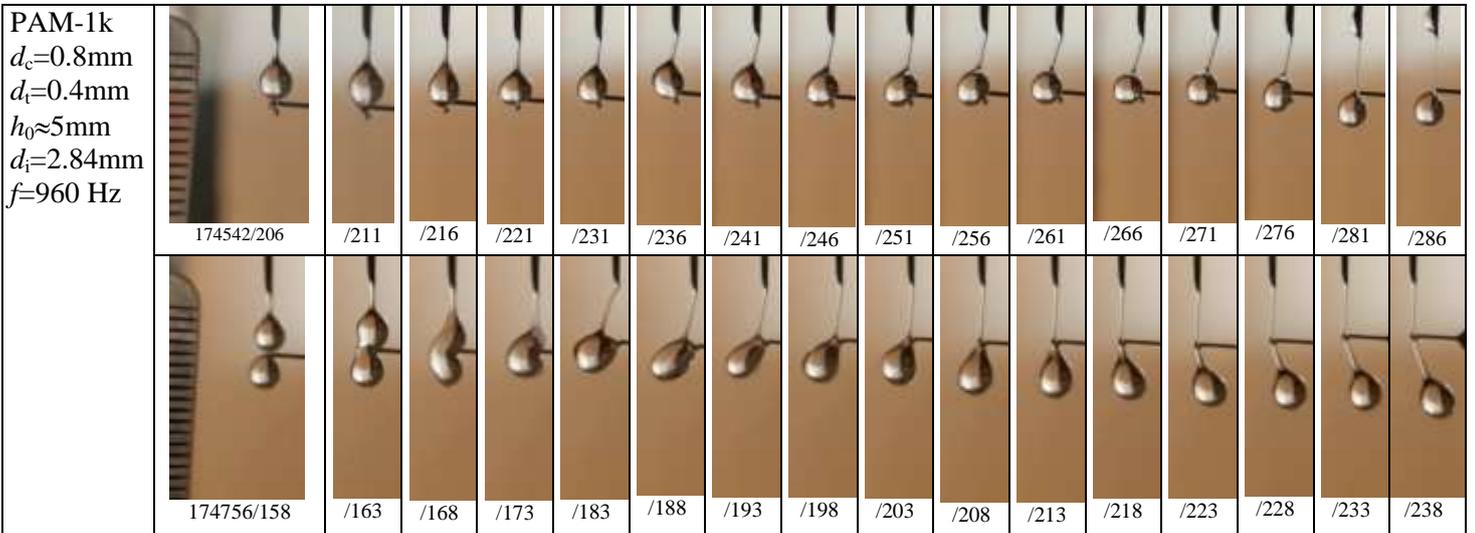

PAM-1k
$d_c$=0.8mm
$d_t$=0.4mm
$h_0$≈5mm
$d_i$=2.84mm
$f$=960 Hz

174542/206 /211 /216 /221 /231 /236 /241 /246 /251 /256 /261 /266 /271 /276 /281 /286

174756/158 /163 /168 /173 /183 /188 /193 /198 /203 /208 /213 /218 /223 /228 /233 /238

Fig. 14. An example of the capture of a PAM-1k drop by an obstacle and the detachment of the captured drop by the impact of the next drop with sagging at $d_c$=0.8mm, $d_t$=0.4mm, $h_0$≈5mm. The ruler on frame 74756/158 shows the scale of the image

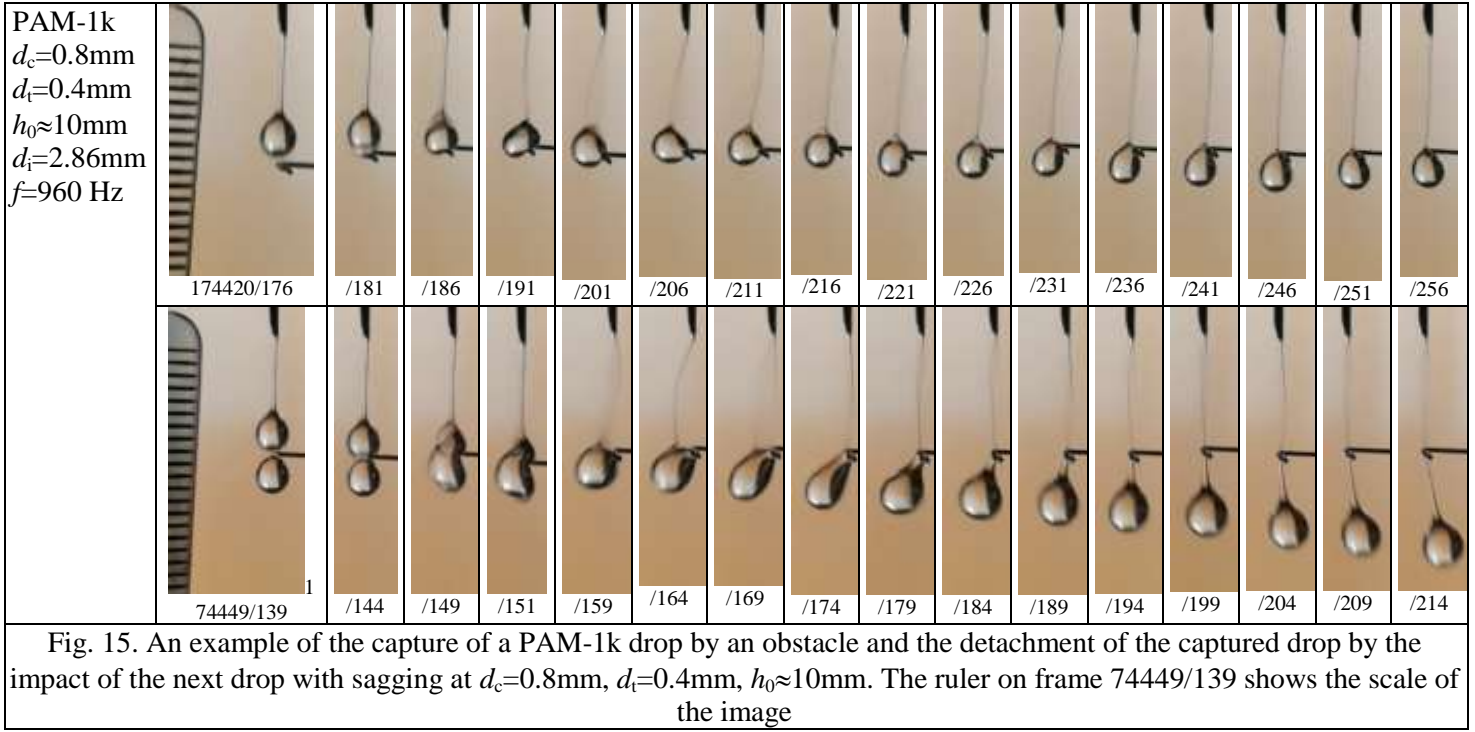

PAM-1k
$d_c$=0.8mm
$d_t$=0.4mm
$h_0$≈10mm
$d_i$=2.86mm
$f$=960 Hz

Fig. 15. An example of the capture of a PAM-1k drop by an obstacle and the detachment of the captured drop by the impact of the next drop with sagging at $d_c$=0.8mm, $d_t$=0.4mm, $h_0$≈10mm. The ruler on frame 74449/139 shows the scale of the image

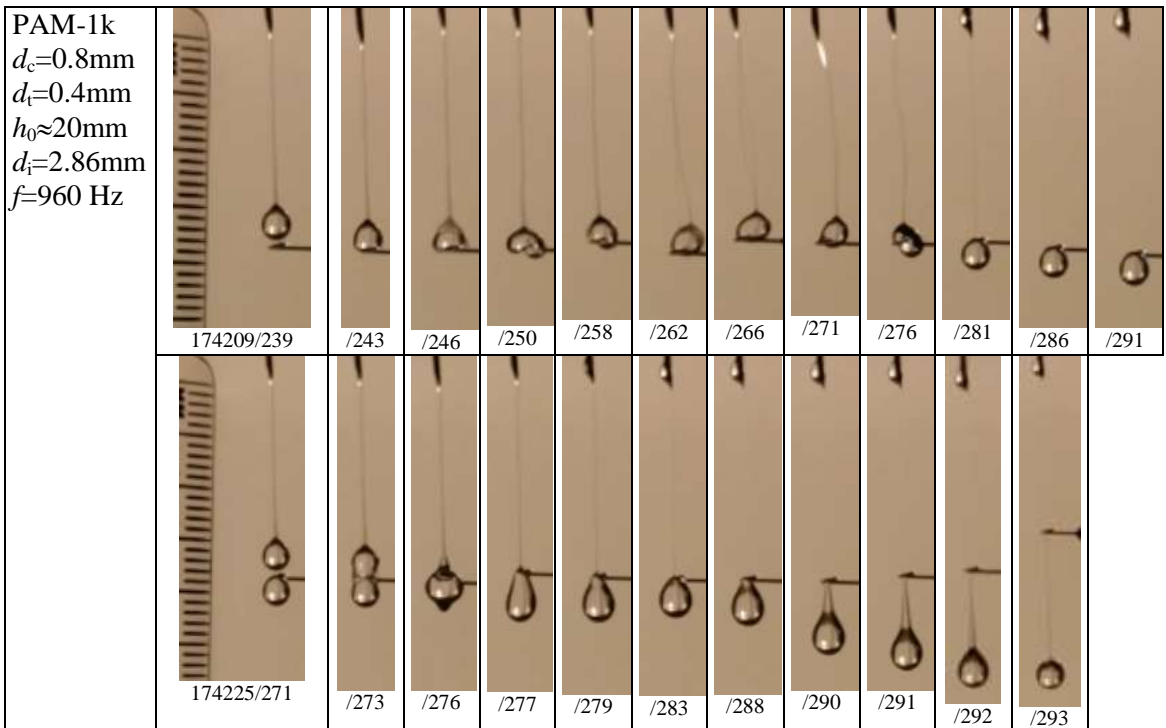

PAM-1k
$d_c$=0.8mm
$d_t$=0.4mm
$h_0$≈20mm
$d_i$=2.86mm
$f$=960 Hz

Fig. 16. An example of the capture of a PAM-1k drop by an obstacle, rebound and detachment of the captured drop by the impact of the next drop at $d_c$=0.8mm, $d_t$=0.4mm, $h_0$≈20mm. The ruler on frame 174225/271 shows the scale of the image

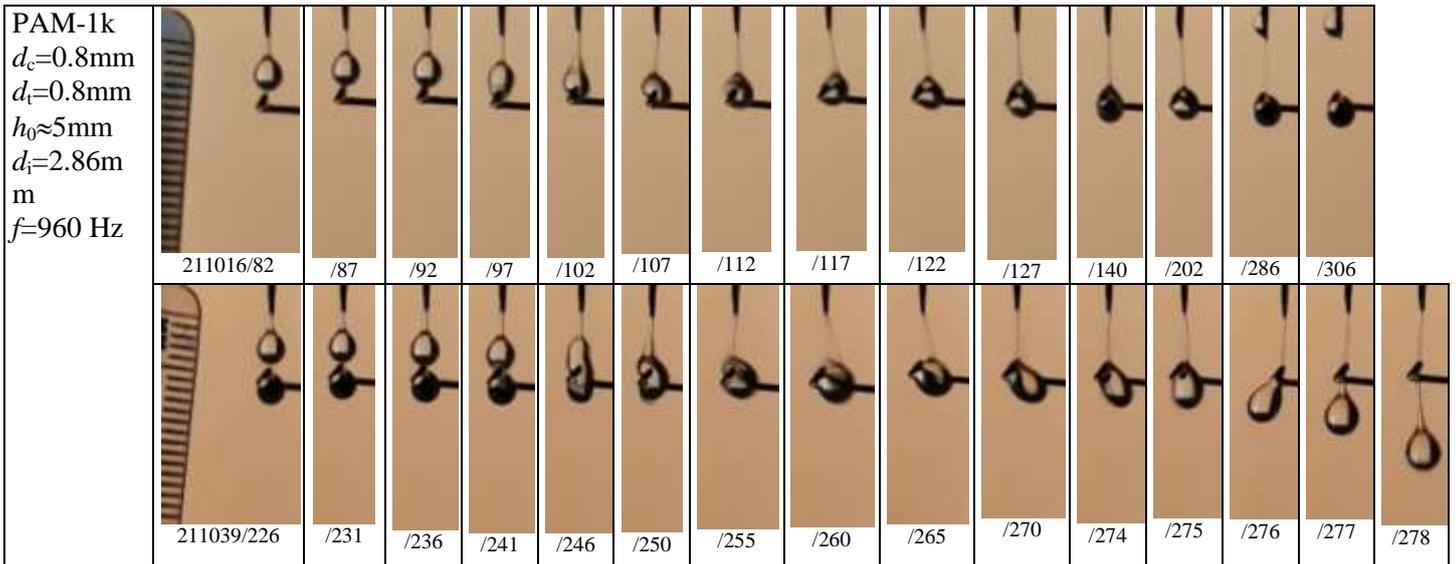

PAM-1k
$d_c$=0.8mm
$d_t$=0.8mm
$h_0$≈5mm
$d_i$=2.86mm
$f$=960 Hz

211016/82  /87  /92  /97  /102  /107  /112  /117  /122  /127  /140  /202  /286  /306

211039/226  /231  /236  /241  /246  /250  /255  /260  /265  /270  /274  /275  /276  /277  /278

Fig. 17. An example of the capture of a PAM-1k drop by an obstacle, and the detachment of the captured drop by the impact of the next drop with sagging at $d_c$=0.8mm, $d_t$=0.8mm, $h_0$≈5mm. The ruler on frame 211039/226 shows the scale of the image

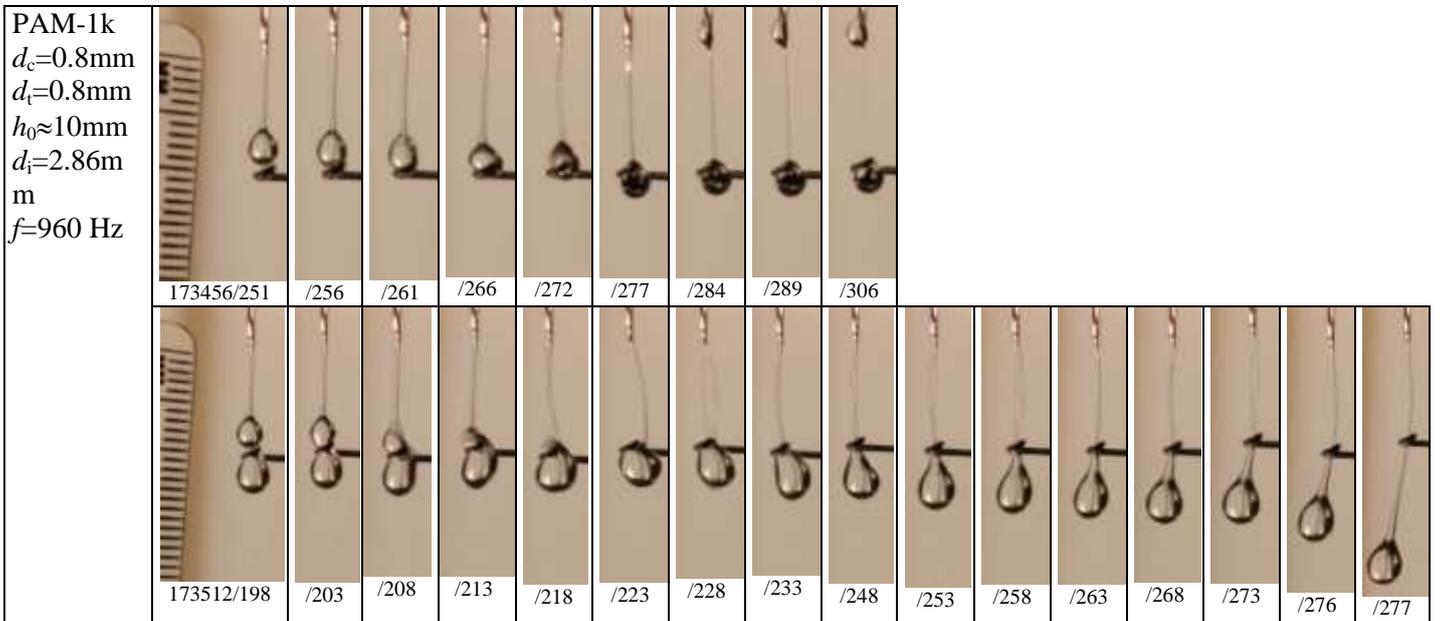

PAM-1k
$d_c$=0.8mm
$d_t$=0.8mm
$h_0$≈10mm
$d_i$=2.86mm
$f$=960 Hz

173456/251  /256  /261  /266  /272  /277  /284  /289  /306

173512/198  /203  /208  /213  /218  /223  /228  /233  /248  /253  /258  /263  /268  /273  /276  /277

Fig. 18. An example of the capture of a PAM-1k drop by an obstacle, and the detachment of the captured drop by the impact of the next drop with sagging at $d_c$=0.8mm, $d_t$=0.8mm, $h_0$≈10mmThe ruler on frame 173512/198 shows the scale of the image

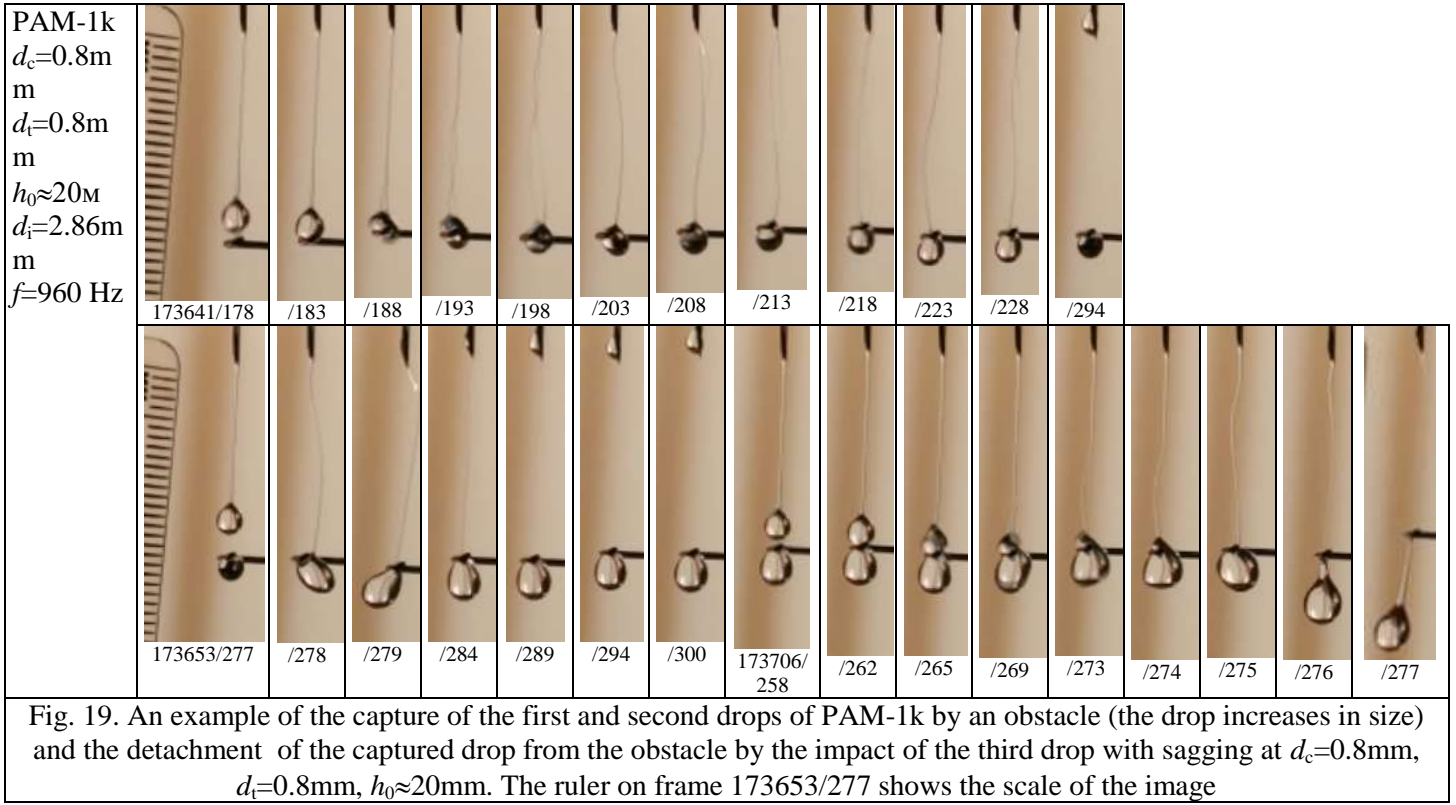

Fig. 19. An example of the capture of the first and second drops of PAM-1k by an obstacle (the drop increases in size) and the detachment of the captured drop from the obstacle by the impact of the third drop with sagging at $d_c$=0.8mm, $d_t$=0.8mm, $h_0$≈20mm. The ruler on frame 173653/277 shows the scale of the image

*3.2 Numerical simulations*

*Setting of the mathematical problem.* In this part of the paper, we consider a mathematical model of the problem of the transverse central collision a water drop with a diameter $d_i$ =0.5 mm on a solid thread of cylindrical shape with a diameter $d_t$ =0.1 mm. In this model, gravity is excluded from consideration. The calculate region is a rectangle with sides of 2 mm and 5 mm. In the center of the calculate region there is thread with cylindrical cross section (Fig. 20). At the initial moment, the drop is located at the distance of its radius from the surface of the thread and has a velocity $\mathbf{u}(u_1, u_2, \mathrm{t}=0)=\mathbf{v}_i$. It is assumed that the "water-air" interface is intentionally slightly blurred at the initial moment (Fig. 20).

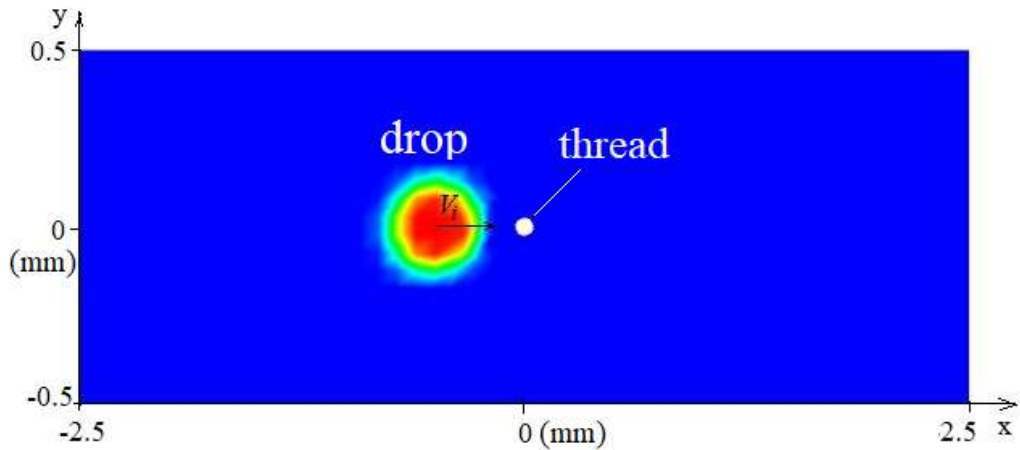

Fig. 20. Calculation region and the location of drop at the initial moment

The range of the initial velocities of drop $v_i$ from 0.1 to 10 m/s is considered. The drop is located in an airstream moving at a speed of 0.1 m/s in the direction of the horizontal x-axis. The following boundary conditions were set: constant velocity at the input of the calculated area (x=-2.5 mm), pressure at the output (x=2.5 mm), and flow symmetry conditions at the horizontal boundaries (y=±0.5 mm). On the interface of the gas - liquid interface, the condition of equilibrium of surface forces and pressure is set. The wetting angle on the thread was set to 90[0].

The mathematical model is based on solving the system of two-dimensional Navier-Stokes equations for a two-phase gas-liquid system in the approximation of the "mixture" model [4] can be written in the form:

$$\frac{\partial u_1}{\partial x} + \frac{\partial u_2}{\partial y} = 0 \tag{1}$$

$$\frac{d(\rho u_1)}{dt} = -\frac{\partial p}{\partial x} + \frac{\partial}{\partial x}\left(\mu \frac{\partial u_1}{\partial x}\right) + \frac{\partial}{\partial y}\left(\mu \frac{\partial u_2}{\partial y}\right) + F_1 \tag{2}$$

$$\frac{d(\rho u_2)}{dt} = -\frac{\partial p}{\partial y} + \frac{\partial}{\partial x}\left(\mu \frac{\partial u_1}{\partial x}\right) + \frac{\partial}{\partial y}\left(\mu \frac{\partial u_2}{\partial y}\right) + F_2 \tag{3}$$

where $df/dt = \partial f/\partial t + u_1 \partial f/\partial x_1 + u_2 \partial f/\partial y$ is the substantial derivative; x, y are Cartesian coordinates; $u_1$, $u_2$ are components of velocity vector $\mathbf{u}(u_1, u_2)$, t is the time, p is the pressure; $\rho$ is density; $\mu$ is the dynamic viscosity coefficients; $F_1$, $F_2$ are components of external force $\mathbf{F}(F_1, F_2)$, operating in a narrow zone along the air-liquid interface.

To describe the two-phase air-liquid system, we used the system of equations (1-3) with one equation for momentum transfer under the assumption of a "mixture" model [4] with averaged velocities $\mathbf{u} = \varepsilon \mathbf{u}_{air} + (1-\varepsilon)\mathbf{u}_{liquid}$, density $\rho = \varepsilon \rho_{air} + (1-\varepsilon)\rho_{liquid}$, and viscosity $\mu = \varepsilon \mu_{air} + (1-\varepsilon)\mu_{liquid}$, where the values with the index air refer to air, and with the index liquid refer to liquid. The volume fraction of liquid $\varepsilon$ ($0 < \varepsilon < 1$) was determined from the solution of the transfer equation: $\partial \varepsilon / \partial t + \partial \varepsilon u_1 / \partial x + \partial \varepsilon u_2 / \partial y = 0$.

The boundary conditions at the air–liquid interface were determined from the equilibrium condition of surface forces and pressure [4]:

$$(p_1 - p_2 + \sigma k)n_i = (\tau_{1ij} - \tau_{2ij})n_j + \partial \sigma / \partial x_i \tag{4}$$

where $\tau_{\alpha ij} = \mu_\alpha \left(\frac{\partial u_i}{\partial x_j} + \frac{\partial u_j}{\partial x_i}\right)_\alpha$ is the viscous stress tensor (i =1,2; j =1,2; $x_1$ = x, $x_2$ = y); index $\alpha$ denotes: $\alpha$=1 – liquid, $\alpha$=2 – air; $\sigma$ is the surface tension coefficient, which depends on the properties of the liquids, the wetting angle, and in common case can be a function of temperature, impurity concentration, and coordinates; $p_1$, $p_2$ are the fluid and air pressures; $\kappa = 1/R_1 + 1/R_2$ is the surface curvature where $R_1, R_2$ are the radii of curvature for liquid and air; $\mathbf{n}(n_1, n_2)$ is the unit normal vector directed into the second fluid.

The conditions for the absence of friction are set at the boundaries of the computational domain and condition (5) at the interface of the two-phase liquid-air system. The modeling of the change in the shape of the air-liquid interface was performed using the model of liquid volumes (VOF - Volume Of Fluid method). The interface was determined by the VOF method with increased resolution and taking into account surface forces by the CSF method (Continuum Surface Force) [5]. The CSF method allows one to remove the singularity in the case of turning the radius of curvature of the interface surface to zero and increase the accuracy of the calculations [5]. When solving the problem, condition (4) at the interphase boundary was taken into account through an additional local bulk force $\mathbf{F}$ on the right side of the momentum transfer equation (2 - 3). The force $\mathbf{F}$ acts only in a very narrow area, enclosed along an interface curve $l$ of width $\Delta h$, where $l$ is the curve which shows the exact location of the air-water interface in the 2D calculation area. When $\Delta h$ tends to zero, $\mathbf{F}$ force can be written as $\mathbf{F}(l_s) = \sigma \kappa(l_s) \mathbf{n}(l_s)$ at each point $l_s$ of the interface curve $l$, where $\kappa(l_s)$ is curvature of curve $l$ at point $l_s$, $\mathbf{n}(l_s)$ is the normal at point $l_s$ of interface curve $l$ [5].

For numerical solution of the system of equations (1-4), the conservative method of control volumes with the approximation of spatial derivatives of the second order and first order in time was used [6]. The accuracy of defining the interface is limited by the size of the grid cells and the solution methods, so a detailed dynamic grid was used on both sides of the interface. The additional details, validation results of the mathematical model and the examples of its using are given in papers [7-9].

*Results of numerical simulations.* The results of numerical simulation of the problem of flow around a thread ($d_t$=0.1 mm) by a water drop ($d_i$=0.5 mm) showed that a single thread is able to delay the penetration of a drop at drop velocities of less than 1 m/s. For example, the drop almost completely passes through the thread at a drop speed equal 1 m/s, but the tension forces hold it at thread, and it sticking to the thread making oscillatory movements near the thread (Fig. 21). Under certain conditions, the drop sticking to the thread can make a precessional rotation around the thread. The shapes and positions of the droplet at different consecutive time points are shown in Fig. 21 ($v_i$ = 0.1 m/s) and Fig. 22 ($v_i$ = 10 m/s). When the drop flows around the thread at a speed of more than 1 m / s, the drop is not delayed by the thread, and the drop torn by the thread into two parts, which are then combined into one drop, which continues to move behind the thread, as shown in Fig. 22.

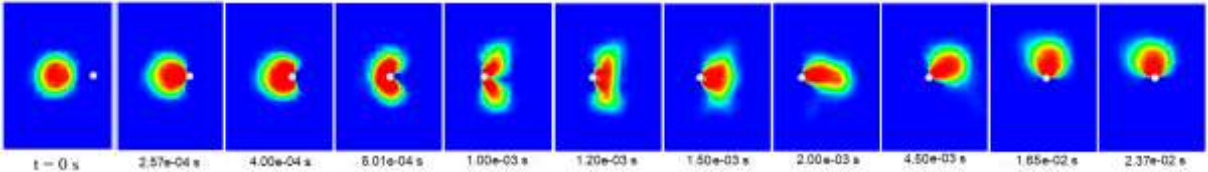

Fig. 21. The shape and position of the drop ($d_i$=0.5 mm) at different moments of time during flow around the thread ($d_t$=0.1 mm), the drop moves from left to right at rate of $v_i$=1 m/s

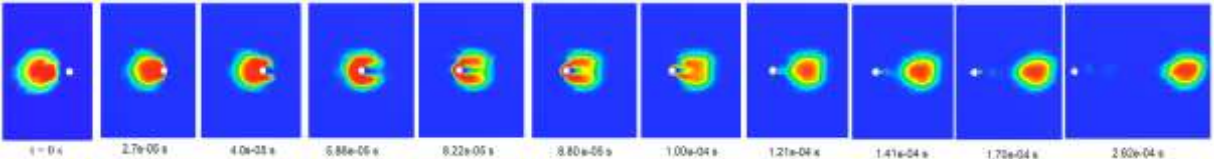

Fig. 22. The shape and position of the drop ($d_i$=0.5 mm) at different moments of time during flow around the thread ($d_t$=0.1 mm), the drop moves from left to right at rate of $v_i$=10 m/s

## 4. Conclusions

Observations have shown that various scenarios of the interaction of a falling drop with a thin cylinder are possible, which depend on the kinematics of the impact and the rheological properties of the liquid. It has been established that the elasticity of the liquid, which increases with the concentration of the polymer, promotes the capture of droplets by a thin cylinder. Detachment of the drop in this case is provided by additional impacts of the following drops. The experiments also revealed the phenomenon of droplet rebound from a thin cylinder, and it is observed both for water and polymer liquids. Earlier, the rebound of droplets was observed when a droplet collided with a flat surface. In addition, it was found in experiments that under certain circumstances a drop breaks down into two, and the further fate of these drops can be different. The results of numerical simulation have shown the presence of different modes of flow around a thin thread by a water drop, which are in good agreement with the experimental data. At drop rates of less than 1 m/s, the thread ($d_t$=0.1 mm) is able to delay the penetration of a water drop ($d_i$=0.5 mm), that is, the drop is held near the thread due to surface tension forces.

## Acknowledgments

The work was supported by grant RFBR 20-04-60128.